\documentclass{natureprintstyle}

\spacing{1}
\spacing{2}
\spacing{1}

\usepackage{amsmath,amssymb}
\usepackage[applemac]{inputenc}
\usepackage{graphicx}
\usepackage{color}
\usepackage{bm}
\usepackage{longtable}
\usepackage{amssymb}
\usepackage{rotating}
\usepackage{hyperref}

\newcommand{\diamonds}{\textsc{D\large{iamonds}}}
\usepackage{hyperref}

\let\citep\cite
\let\citet\cite

\title{\sffamily Spin alignment of stars in old open clusters}

\author{\sffamily Enrico Corsaro$^{1,2,3,4}$, Yueh-Ning Lee$^1$, Rafael A. Garc\'{i}a$^{1}$, Patrick Hennebelle$^1$, Savita Mathur$^{5}$, Paul G. Beck$^1$, Stephane Mathis$^1$, Dennis Stello$^{6,7}$ \& J\'er\^{o}me Bouvier$^8$}

\begin{document}
\maketitle
\let\thefootnote\relax\footnote{

\begin{affiliations}
\item Laboratoire AIM Paris-Saclay, CEA/DRF --- CNRS --- Universit\'e Paris Diderot, IRFU/SAp Centre de Saclay, F-91191 Gif-sur-Yvette Cedex, France
\item Instituto de Astrof\'{i}sica de Canarias, E-38200 La Laguna, Tenerife, Spain
\item Departamento de Astrof\'{i}sica, Universidad de La Laguna, E-38205 La Laguna, Tenerife, Spain
\item INAF - Osservatorio Astrofisico di Catania, Via S. Sofia 78, I-95123 Catania, Italy
\item Space Science Institute, 4750 Walnut street Suite 205, Boulder, CO 80301, USA
\item Sydney Institute for Astronomy (SIfA), School of Physics, University of Sydney, Sydney, New South Wales 2006, Australia
\item Stellar Astrophysics Centre, Department of Physics and Astronomy, Aarhus University, Ny Munkegade 120, DK-8000 Aarhus C, Denmark
\item Universit\'e Grenoble Alpes, IPAG, F-38000 Grenoble, France; CNRS, IPAG, F-38000 Grenoble, France
\end{affiliations}
}

\vspace{-3.5mm}
\begin{abstract}
\sffamily Stellar clusters form by gravitational collapse of turbulent molecular clouds, with up to several thousand stars per cluster\cite{Lee12}. They are thought to be the birthplace of most stars and therefore play an important role in our understanding of star formation, a fundamental problem in astrophysics\cite{Lada03,Longmore14}. The initial conditions of the molecular cloud establish its dynamical history until the stellar cluster is born. However, the evolution of the cloud's angular momentum during cluster formation is not well understood\cite{McKee07}. Current observations have suggested that turbulence scrambles the angular momentum of the cluster-forming cloud, preventing spin alignment amongst stars within a cluster\cite{JJ10}. Here we use asteroseismology\cite{Gizon03,Beck12,Huber13} to measure the inclination angles of spin axes in 48 stars from the two old open clusters NGC~6791 and NGC~6819. The stars within each cluster show strong alignment. Three-dimensional hydrodynamical simulations of proto-cluster formation show that at least 50\% of the initial proto-cluster kinetic energy has to be rotational in order to obtain strong stellar-spin alignment within a cluster. Our result indicates that the global angular momentum of the cluster-forming clouds was efficiently transferred to each star and that its imprint has survived after several gigayears since the clusters formed.
\end{abstract}

\sffamily
About half of the overall star formation in the Milky Way is occurring in the 24 most massive giant molecular clouds\cite{Lee12}. The star forming regions are obscured by dust, hence direct observations are limited to the infrared and radio bands\cite{Lada03,Longmore14}. However, open clusters can be studied in a broad range of wavelengths because they contain small amounts of interstellar gas and dust. The great advantage of studying stars in a cluster --- as opposed to field stars that often originate from dissolved small stellar systems --- is that they can preserve the signature of the initial conditions of the progenitor molecular cloud.

It is believed that molecular clouds satisfying the Jeans instability undergo gravitational fragmentation in which the internal motions are strongly influenced by turbulence\cite{McKee07,Urquhart14}. This suggests that the angular momentum from the progenitor cloud cannot leave any significant imprint of its action on the stars born in the cluster. However, if the stars inherit the physical properties of the molecular cloud, they should to some extent reflect its average angular momentum. To investigate the angular momentum imprint, requires measurements of the space orientation of the stellar-spin axis. Previous analyses conducted on young open clusters did not find evidence of stellar-spin alignment\cite{JJ10}.

Asteroseismology, the study of stellar oscillations, has proven to be a powerful tool to obtain model-independent information on the inclination angle of the stellar angular momentum vector, especially for red giant stars\cite{Ballot06,Benomar15,Beck12,Huber13}. Red giants are typically low- and intermediate-mass stars that have evolved off the main sequence of the stellar evolution. Most red giants oscillate and their oscillations can be analyzed through a Fourier frequency spectrum of their light curve. The spectrum of a red giant contains a comb-like structure of tens and sometimes more than a hundred radial and non-radial oscillation modes, most of which are mixed modes originating by the coupling between acoustic and gravity modes\cite{Bedding11}. Each oscillation mode is identified by an angular degree {\it l}, which gives rise to a multiplet of ($2 l + 1$) different components through the degeneracy lifted by the stellar rotation\cite{Gizon03}. Each rotationally split component is in turn identified by an azimuthal number, $m \leq | l |$. The dipolar ($l = 1$) mixed modes are the most suited for measuring the orientation of the spin axis in red giants\cite{Aerts10}.

We have investigated 48 oscillating red giant stars, with typical masses within the range $\sim1.1$-$1.7$\,$M_{\odot}$ ($M_{\odot}$ being the mass of the Sun), that belong to the open clusters NGC~6791 and NGC~6819\cite{Basu11,Stello11,Corsaro12}. The most relevant physical properties of the two open clusters are outlined in Table~1. Both clusters are old, with NGC~6791 being one of the oldest known in our Galaxy\cite{Brogaard12}, which implies that the initial molecular clouds were massive enough to ensure that the cluster evaporation time --- the time it takes for all the members of a cluster to be ejected by internal stellar encounters --- is well beyond the gigayear (Gyr) time scale. We used four years of time-series photometry obtained by NASA's Kepler mission. We measure the asteroseismic properties\cite{Corsaro14,Corsaro15} of a total of about 380 rotationally split dipolar mixed modes --- identified from a set of more than 3900 oscillation modes --- and we use them to measure the spin-inclination angles (see Methods and Supplementary Fig. 1 for an example of the fits to the oscillation modes). 

The distributions of the spin inclinations (Fig. 1), show that about $70$\% of the stars in each cluster have a strong level of alignment, with a low to mid angle for NGC~6791 ($\theta \approx 30^{\circ}$ and an alignment coefficient $\alpha \simeq 0.77$) and a low angle for NGC 6819 ($\theta \approx 20^{\circ}$, $\alpha \simeq 0.74$), close to a pole-on configuration (see Methods for more details, and Supplementary Tables 1 and 2 for a list of all the results). We also notice a clear cut-off for higher angles (mid to high) in both clusters, where instead a larger number of stars would be expected if the spin vectors were uniformly (randomly) distributed in three dimensions. The binary stars identified in NGC 6819 appear to follow a similar alignment trend to that of the single star members of the cluster. The probability that the observed levels of alignment are the result of a random distribution is less than $1$ in $10^9$ for NGC~6791, and $1$ in $10^7$ for NGC~6819. Conversely, the distribution obtained for an independent sample of $36$ field red giants (not members of any cluster) shows no significant stellar-spin alignment ($\alpha \simeq 0.38$), even when considering subsamples of stars in the same evolutionary stage ($\alpha \simeq 0.41$ for shell-H-burning and $\alpha \simeq 0.35$ for core-He-burning stars, respectively, see Methods and Supplementary Table~5). In addition our reanalysis of an independent sample of main-sequence stars in NGC 6819 hinting at high spin-inclination angles from their rotational periods and projected equatorial velocities\cite{Meibom15}, reveals that their period measurements are not reliable and hence not in conflict with our low-inclination results (see Methods and Supplementary Fig.~2 and 3 for more details). In Fig.~2 we provide the spatial positions of the cluster red giants within fields of view approximately corresponding to the observed size of each cluster, as reported in Table~1. The stars appear to sample the entire field of each cluster, with a tangential distance from star to star varying from $\approx 0.1$\,parsec (pc) near the center up to a few pc in the peripheral regions. This demonstrates that the spin alignment is observed across the entire clusters.

N-body simulations coupled with observations of old open clusters, aimed at reproducing their dynamical evolution, show that the stellar angular momentum can have an impact on colliding stars and on the orbital configurations of multiple stellar systems\cite{Geller13}. The orbital parameters of eccentricity, inclinations, and periods of multiple stellar systems are mostly influenced by tidal forces\cite{Hut81}. For individual stars the angular momentum typically evolves as a spin down over time\cite{Meibom15} through either mass loss by stellar winds, magnetic braking, or tidal friction if the star is captured to form a binary. Given the average distances among the star members of an open cluster, the effect of gravitational N-body interactions on producing any significant spin alignment is negligible even for timescales of several Gyr. This is especially true for tidal forces, because their strength is a function of the inverse third power of the distance between two stars. Furthermore, using the galactic latitudes and distance moduli reported in Table 1, the heights of the two clusters from the galactic plane are $\approx 800\,$pc and $\approx 350\,$pc for NGC~6791 and NGC~6819, respectively, showing that they are located far from the most crowded regions that constitute the inner galactic disk. The position and long survival time of the two clusters therefore suggests that they  experienced significantly less disruptive encounters with giant molecular clouds than other open clusters located inside the disk\cite{vandenbergh80}. This means that the strong spin alignment observed in the red giants of our sample is very likely to be originating from the formation epoch of the cluster, thus preserving the signature of the early dynamical processes characterizing the progenitor molecular cloud.

For exploring this scenario we have performed three-dimensional hydrodynamical simulations of a collapsing molecular cloud leading to the formation of a proto-cluster under the action of gravitational potential\cite{Lee16} (see Methods for further explanations). Figure 3a shows that when considering only a turbulent velocity field, the distribution of the resulting spin inclinations resembles that of a uniform orientation in three dimensions. By introducing a global rotation as an additional initial condition and imposing a ratio of rotational kinetic energy over turbulent kinetic energy of $E_\mathrm{rot}/E_\mathrm{tur} \simeq 1$, the spin alignment produced in stars forming with masses greater than $0.7\,M_{\odot}$ becomes comparable to that found in our observations, with $\alpha \simeq 0.6$ (Fig. 3b and c). When $E_\mathrm{rot}/E_\mathrm{tur} < 1$ instead, the spin alignment is only marginal (see Supplementary Table~3 for all the values we investigated). From our simulations we also observe that for stellar masses below $0.7\,M_{\odot}$ the turbulent motions still dominate over the effect of rotation even in the case $E_\mathrm{rot}/E_\mathrm{tur} \simeq 1$, thus preventing any spin alignment among stars of this low mass range. This suggests that when not enough mass from the molecular cloud is accreted into individual pre-stellar cores, the information from the cloud's average angular momentum is lost in the stellar-spin fluctuations induced by turbulence at the scales of the forming star. A strong component of the cloud's rotational kinetic energy, at least comparable to that of the turbulence, can therefore be responsible of efficiently aligning the spin axes within the stellar members of a cluster because the degree of alignment reflects the importance of the cloud's average angular momentum. The two open clusters NGC~6791 and NGC~6819 could have originated through a formation process involving a compact collapsing molecular cloud giving rise to a rotating proto-cluster (Fig.~3c).  During the cluster formation, stars with masses at least that of our Sun are more likely to inherit a significant fraction of the cloud's average angular momentum with respect to stars having masses below $0.7\,M_{\odot}$. By measuring stellar spin inclinations for solar-mass stars in open clusters we can therefore constrain the initial energy budget of the progenitor molecular cloud, its global rotation, as well as the efficiency by which the cloud's average angular momentum is transferred to the individual stellar members of the clusters. This result allows us to explore and reconstruct the dynamical evolution, structure, and geometry of galactic star forming regions that have formed stellar clusters back in times comparable to the age of our Universe.

\begin{addendum}
\item [Acknowledgements] E.C. is funded by the European Community's Seventh Framework Programme (FP7/2007-2013) under grant agreement n$^\circ$\,312844 (SPACEINN) and by the European Union's Horizon 2020 research and innovation programme under the Marie Sklodowska-Curie grant agreement n$^\circ$\,664931. Y.-N.L. and P.H. acknowledge funding by the European Research Council under the European Community's Seventh Framework Programme (FP7/2007-2013 grant agreement n$^\circ$\,306483) and the HPC resources of CINES under the allocation x2014047023 made by GENCI (Grand Equipement National de Calcul Intensif). R.A.G. received funding from the CNES GOLF and PLATO grants at CEA. R.A.G. and P.G.B. received funding from the ANR (Agence Nationale de la Recherche, France) program IDEE (n$^\circ$\,ANR-12-BS05-0008) ``Interaction Des \'Etoiles et des Exoplan\`etes''. Sa.M. acknowledges support from the NASA grant NNX12AE17G. St.M. acknowledges funding by the European Research Council through ERC grant SPIRE n$^\circ$\,647383. D.S. is the recipient of an Australian Research Council Future Fellowship (project n$^\circ$\,FT140100147). J.B. acknowledges financial support from grant ANR 2011 Blanc SIMI5-6 020 ``Toupies: Towards understanding the spin evolution of stars''. This work has received funding from the CNES grants at CEA. All the light curves used in this paper were obtained from the Mikulski Archive for Space Telescopes (MAST). STScI is operated by the Association of Universities for Research in Astronomy, Inc., under NASA contract NAS5-26555. Support for MAST for non-HST data is provided by the NASA Office of Space Science via grant NNX09AF08G and by other grants and contracts. UKIRT is supported by NASA and operated under an agreement among the University of Hawaii, the University of Arizona, and Lockheed Martin Advanced Technology Center; operations are enabled through the cooperation of the East Asian Observatory. We thank David Salabert for the preparation of the website containing the source data used in this work. 

\item[Author contributions] E.C. performed the fits of the background in the power spectra, identified the oscillation modes, measured the mode parameters and the inclination angles for all the stars in the sample, and interpreted the results. Y.-N.L. performed the hydrodynamical simulations of the proto-cluster formation and for the significance of the stellar-spin alignment, and contributed in interpreting the initial conditions in the molecular cloud. R.A.G. prepared the datasets calibrated for the asteroseismic analysis, contributed in discussing the analysis method and the observational results, and reanalyzed the independent sample of stars observed in NGC 6819. P.H. contributed in the computation of the hydrodynamical simulations and in the interpretation of observational results and of the initial conditions in the molecular cloud. S.Mathur provided input guesses for the background properties in the power spectra of all the stars and contributed in the selection of the control sample. P.G.B. contributed in discussing the data analysis method and the identification of the oscillation modes. S.Mathis contributed in discussing the N-body interactions among stars in open clusters and in quantifying the tidal effects in binary stars. D.S. provided spatial positions for the entire population of red giants identified in the field of the two clusters and contributed in discussing the observational results and the data analysis method. J.B. provided theoretical and observational insights on the effect and evolution of angular momentum in stellar clusters. All authors commented on the manuscript.

\item[Author information]Supplementary information is available for this paper. Reprints and permissions information is available at \href{www.nature.com/reprints}{www.nature.com/reprints}.
Correspondence and requests for materials should be addressed to E.C. (e-mail: \href{mailto:enrico.corsaro@oact.inaf.it}{enrico.corsaro@oact.inaf.it}).

\item[Competing interests]The authors declare no competing financial interests.

\end{addendum}

\newpage
\begin{table}
\sffamily
\centering                         
\begin{tabular}{l r r}       
\hline
\\[-8pt]
\small \textbf{Open cluster} & \small  \textbf{NGC~6791} & \small  \textbf{NGC~6819}\\[1pt]
\hline
\\[-8pt]
\small Total mass ($M_{\odot}$) & \small  $\sim5000$  & \small  $\sim2600$\\[1pt]
\small Distance modulus $(m-M)_0$ & \small  $13.11 \pm 0.06$  & \small  $11.85 \pm 0.05$\\[1pt]
\small Size (pc) & \small $\sim10$  & \small  $\sim7$\\[1pt]
\small Age (Gyr) & \small  $\sim8.3$  & \small  $\sim2.4$\\[1pt]
\small Galactic coordinates & \small $70.0$ (long.),\,$+10.9$ (lat.) & \small $74.0$ (long.),\,$+8.5$ (lat.)\\[1pt]
\small $\overline{M}_\mathrm{RC}$ ($M_{\odot}$) & \small $1.15 \pm 0.03$  & \small  $1.65 \pm 0.04$\\[1pt]
\small Stars analyzed & \small 25 & \small 23 \\[1pt]
\hline
\end{tabular}
\vspace{-2mm}
\caption{\sffamily \small \textbf{Global properties of NGC~6791 and NGC~6819.} The total masses in units of a solar mass show that both clusters are massive, with NGC~6791 one of the most massive known in the Galaxy\cite{Platais11,Kalirai01}. The size is computed in this study by means of the angular radii\cite{Platais11,Kalirai01} and the distance modulus from asteroseismology\cite{Basu11}. Ages are based on eclipsing binaries and triple systems\cite{Brogaard12,Brewer16}. The galactic coordinates refer to the astronomical epoch J2000. The average mass of a core-He-burning star, $\overline{M}_\mathrm{RC}$, is calculated from asteroseismic scaling relations and it is comparable to that of the shell-H-burning red giants of the same cluster because of the low mass loss\cite{Miglio12}. The total number of stars analyzed refers to the pulsating red giants whose orientation of the spin axis has been measured in this work. The corresponding $1\sigma$ random error is shown for distance modulus and average mass of the stars.}             
\label{tab:properties}
\vspace{-4mm}
\end{table}

\begin{figure}[tb]
\centerline{
\includegraphics[width=0.47\textwidth]{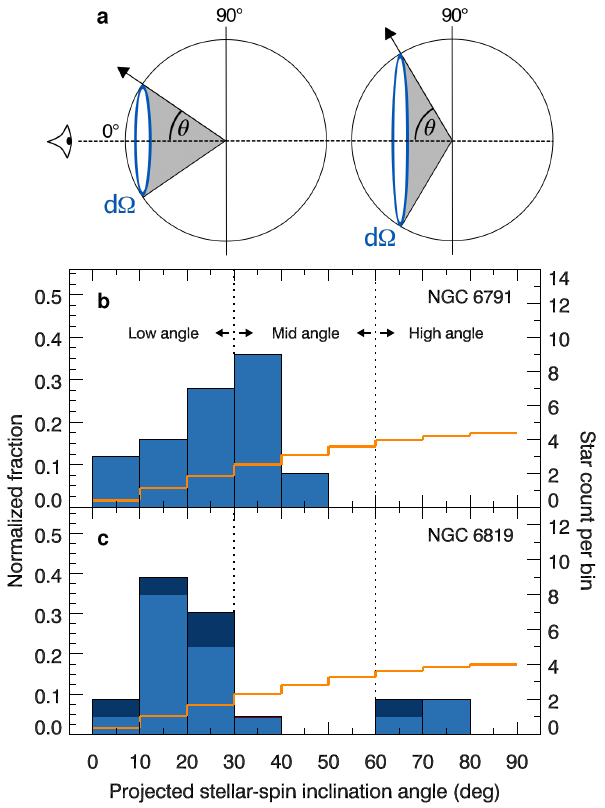}
}
\vspace{-2mm}
\caption{\sffamily \small \textbf{Projected stellar-spin inclinations of the 48 red giants of NGC~6791 and NGC~6819.} {\sffamily \textbf{a}}, The spin vector of the star (arrow) as seen from the line of sight can be oriented along any directions of a cone (gray shaded), with an inclination angle from $\theta = 0^{\circ}$ (pole-on) to $\theta = 90^{\circ}$ (edge-on). The infinitesimal solid angle $d \Omega$ (blue coronal shell) increases by increasing $\theta$. {\sffamily \textbf{b}}, The distribution of $\theta$ from rotationally split $l = 1$ mixed modes for NGC~6791. The orange histogram shows the expected distribution for a three-dimensional uniform orientation of the spin vectors (see Methods). The vertical dotted lines separate the three main configurations of stellar inclination. {\sffamily \textbf{c}}, Same as panel b but for NGC 6819. The darker regions correspond to red giants that are confirmed spectroscopic single-lined binaries\cite{Milliman14}.
}
\vspace{-4mm}
\end{figure}

\begin{figure}[tb]
\centerline{
\includegraphics[width=0.55\textwidth]{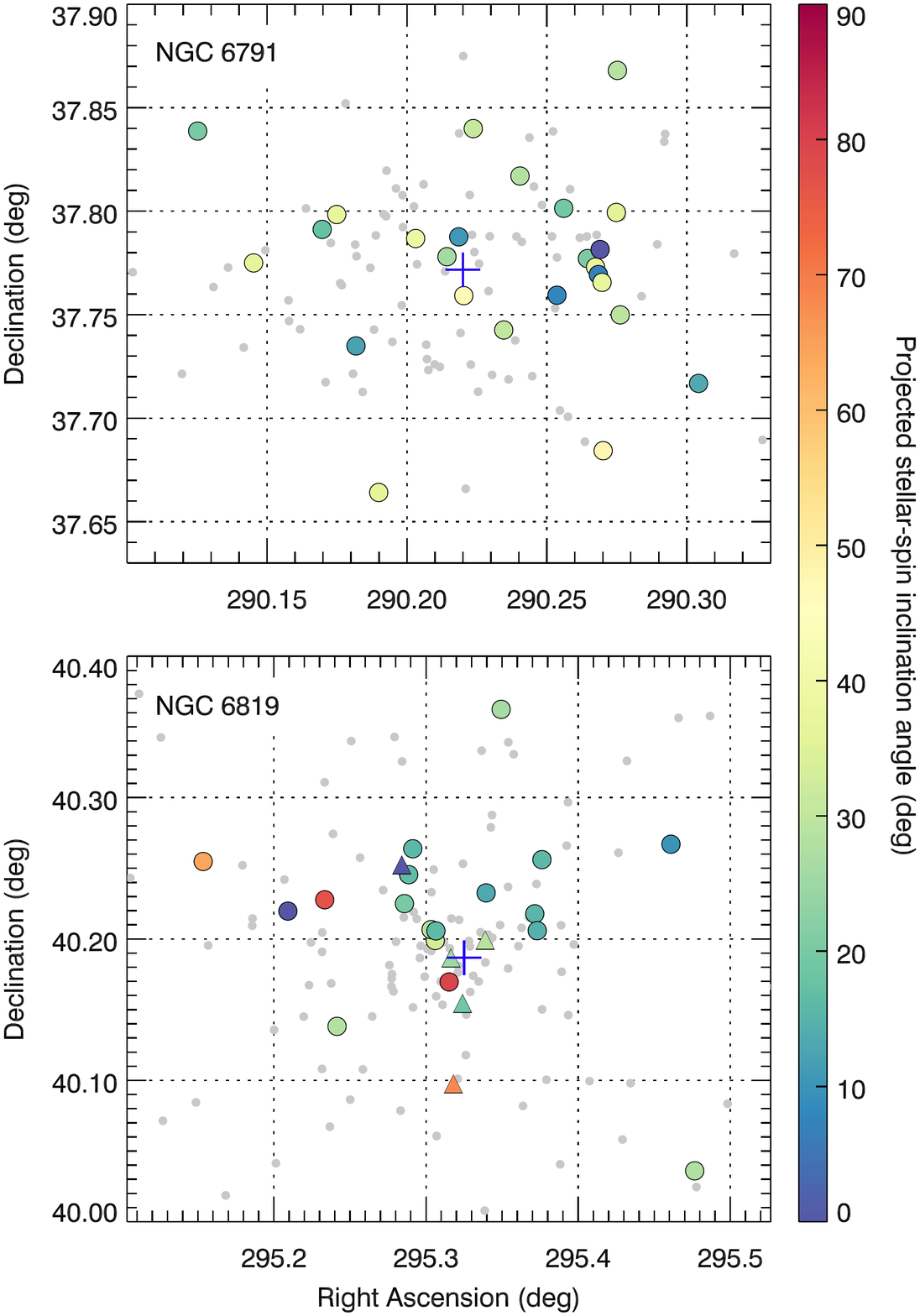}
}
\vspace{-4mm}
\caption{\sffamily \small \textbf{Spatial positions of the 48 red giants of NGC~6791 and NGC~6819.}  The spatial positions of the red giants analyzed here are shown in spherical coordinates (see Supplementary Tables 1 and 2), with projected inclination angles of the stellar-spin axis indicated by a color-coded scale. The group of four stars at high angles in NGC 6819 appears to be localized along a diagonal in the bottom-left region of the diagram. The spectroscopic binaries of NGC 6819 are indicated by triangles. The entire population of red giants identified in the clusters is shown by the gray dots in the background\cite{Stello11}, while the centers of the cluster fields are marked by blue crosses.}
\vspace{-4mm}
\end{figure}

\begin{figure*}
\centerline{
\includegraphics[width=1.0\textwidth]{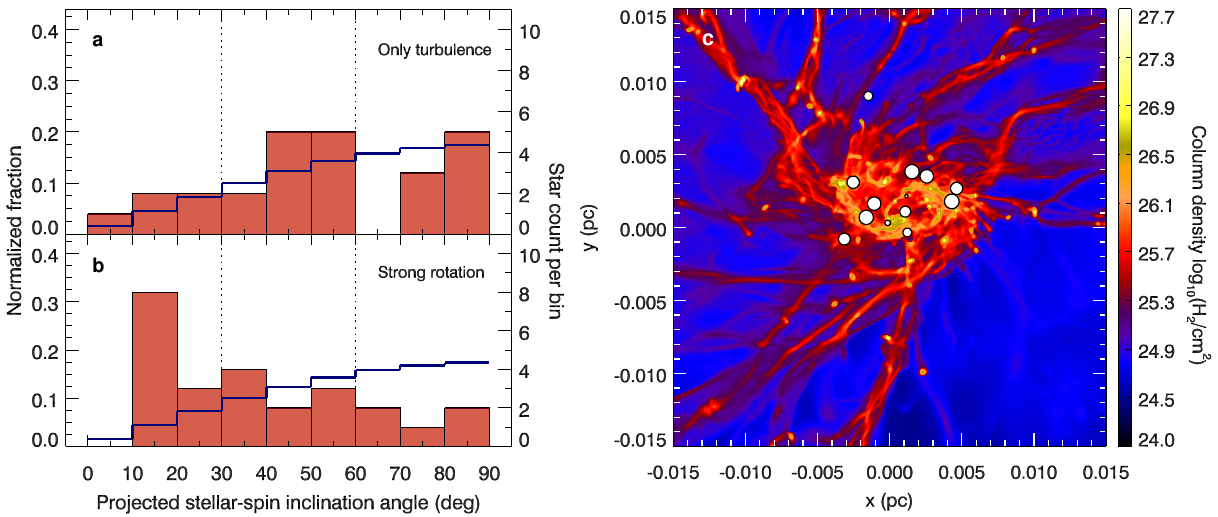}
}
\vspace{-2mm}
\caption{\sffamily \small \textbf{Projected stellar-spin inclinations and proto-cluster formation from hydrodynamical simulations.} {\sffamily \textbf{a}}, The distribution of $\theta$ for 25 stars with $M \geq 0.7$\,$M_{\odot}$ obtained from the simulation without global rotation (Simulation 1 in Supplementary Table 3), where no stellar-spin alignment is detected. The blue histogram represents the expected three-dimensional uniform distribution. {\sffamily \textbf{b}}, Same as panel a but for the simulation with the best level of alignment (Simulation 4). {\sffamily \textbf{c}}, The rotating proto-cluster from Simulation 4 as seen along the axis of best spin alignment $\vec{z}$ (see Methods), with a color-coded column density. White circles represent pre-stellar cores gathering most of the cloud's average angular momentum, with their size proportional to the cosine between their spin axis and $\vec{z}$.}
\vspace{-4mm}
\end{figure*}

\clearpage
\setcounter{page}{1}
\setcounter{figure}{0}
\setcounter{table}{0}

\begin{center}
{\textbf{ \Large \uppercase{Methods}} }
\end{center}

\noindent
{\textbf{Target selection and data preparation.}}  Our red giant stars are selected by having either available estimates of the observed period spacing of mixed dipole modes\cite{Corsaro12} or asymptotic period spacing of dipole gravity modes, $\Delta\Pi_1$, namely the regular spacing in period that characterizes the gravity modes of oscillation\cite{Mosser12,Mosser14}. These period spacings ensure us an unambiguous classification of the evolutionary stage of the stars (either shell-H burning or core-He burning) and represent the starting point for a reliable mode identification of the oscillation peaks in the Fourier spectrum. The selected cluster stars have most precise and largest number of identified stellar oscillation frequencies, which makes them the best suited for the measurement of spin inclination. For NGC 6791 we identify 31 stars but we discard KIC~2435987, KIC~2436097, KIC~2436458, KIC~2437240, KIC~2437589 from the analysis because they are red giant branch stars with a low frequency of maximum power ($\nu_\mathrm{max} \leq 50\,\mu$Hz) and low period spacing ($\Delta\Pi_1 \leq$ 70\,s). For these stars the identification of the rotational multiplets is not reliable because the frequency separation between adjacent dipole mixed modes is less than 10 times the frequency resolution ($\sim$0.008 $\mu$Hz, see below). Finally, KIC~2436593 is removed because of the presence of oscillation power excess from two stars. For NGC 6819 we identify 24 stars but we remove KIC~5024967 because few dipolar mixed modes were found. 

The Kepler light curves used cover the full observing length (up to quarter 17.2, i.e. 4 years) and were optimized for asteroseismic analysis using an inpainting interpolation\cite{Garcia11,Garcia14,Pires14}. We also consider 36 field red giants, obtained by combining already published results for 19 shell-H-burning red giants\cite{Corsaro15} with a new set of 17 core-He-burning red giants. All these field red giants exhibit a signal-to-noise ratio in the power spectra and masses similar to those of the cluster stars, and were observed by Kepler up to quarter 17.2. These field stars constitute our control sample and are selected using the same criterion as for the cluster red giants, hence through the available period spacings\cite{Mosser12,Mosser14,Vrard16}. The 17 core-He-burning red giants represent all the stars available from the latest release of the Kepler catalogue\cite{Mathur16} that satisfy our selection criterion. The large fraction of high spin-inclination angles found for both types of field red giants (see Supplementary Table~5), in agreement with random spin orientation, demonstrates that independently of the evolutionary stage our selection criterion is not introducing any specific bias against high angles.

\noindent
{\textbf{Background signal, oscillation mode properties, and mode identification.}} The first step performed in the analysis of the stellar power spectra is the estimation of the background signal (dashed blue line in Supplementary Fig.~1), namely the signal originating from the stellar granulation, the surface rotation, and the instrumental noise. We have adopted the Bayesian code \diamonds\cite{Corsaro14} and a background model defined by three super-Lorentzian profiles (one for the surface rotation and two for the granulation signal), a flat white noise component, and an additional super-Lorentzian profile accounting for the colored noise at low frequency ($<$ 10 $\mu$Hz), with uniform prior distributions for each free parameter of the background model\cite{Kallinger14,Corsaro15}. The prior distributions were set using first guesses obtained by means of the A2Z pipeline\cite{Mathur10}. 

The second step involves the fitting and identification of the individual oscillation modes, a procedure known as peak bagging analysis\cite{Corsaro14,Corsaro15}. The fitting is done using \diamonds, providing accurate estimates for the properties of mode frequency, linewidth, and amplitude (or height) for each of the oscillation peaks considered. The significance of peaks with a low signal-to-noise ratio (having a height that corresponds to about eight times the background level) is tested by means of Bayesian evidence, and peaks are deemed significant when reaching a detection probability $\geq 99\,$\%. All significant peaks are considered as true oscillation modes and their fitted properties can thus be exploited for further analysis. The mode identification of the acoustic modes ($l = 0, 2, 3$ modes) is done by exploiting the published global asteroseismic parameters of the asymptotic relation for acoustic modes\cite{Tassoul80,Corsaro12}. To identify the dipolar mixed modes in our stars we derive accurate estimates of both $\Delta\Pi_1$ and the coupling factor $q$ of the asymptotic relation of mixed modes\cite{Mosser12}, using a two-dimensional grid-search method to fit the associated asymptotic relation to the power spectra\cite{Corsaro15,Buysschaert16}. We use the published values of the observed period spacing or, when available, $\Delta\Pi_1$ as initial guesses for the fits\cite{Corsaro12,Mosser14}. The mode identification process only relies on the value of $\Delta\Pi_1$, and it is performed in exactly the same way for both shell-H-burning and core-He-burning red giants. The asymptotic relation of mixed modes yields the identification of the $m = 0$ components of each mixed mode multiplet. The rotationally split components of dipolar mixed modes, $m = \pm 1$, can then be easily identified thanks to their nearly symmetrical pattern with respect to the central non-split component, $m=0$. An example result of the complete analysis process for one cluster red giant is provided in Supplementary Fig.~1. The same approach is applied to the control sample.

\noindent
{\textbf{Extraction of projected stellar-spin inclination.}}
The fitting of the projected inclination of the stellar-spin axis is done subsequently to the peak bagging analysis. For this purpose, we first select from four to seven different rotationally-split multiplets of $l = 1$ mixed modes identified from the power spectra of each star. Each multiplet can be described by one to three oscillation peaks ($m$ components), which are fit independently from one another during the peak bagging analysis to allow for a better match to the observations. The multiplets are chosen in a way that they cover at least two different but consecutive radial mode orders. The radial mode orders considered are those containing the highest signal-to-noise ratio peaks, hence the closest to $\nu_\mathrm{max}$. By choosing the highest signal-to-noise peaks we guarantee a better and more reliable constraint of the inclination angle during the fitting process. In addition, the selected multiplets are in the most p-dominated regions of the mixed mode forest, which provides a closer resemblance to the concept of energy equipartition and therefore of the visibility within the $m$ components of the multiplet\cite{Gizon03}. Once the multiplets are chosen, we consider only those $m$ components that correspond to peaks deemed significant during the peak bagging analysis. This prevents us, independently of the signal-to-noise ratio in the power spectra, from using peaks that could constitute just pure noise, otherwise hampering the reliability of the results. 

The oscillation mode properties of frequency, linewidth and amplitude (or height) of each $m$ component of a multiplet are used to build its 3-component fitting model profile\cite{Gizon03,Ballot06,Benomar15}, constructed as the sum of the same peak-profile functions used during the peak bagging analysis. The model profile is constrained by setting its total power density to the value given by the sum of the heights of its individual $m$ components, as measured by the peak bagging analysis. The frequency of each multiplet component $m$ is set to the already available frequency value extracted with the peak bagging analysis. When the central component ($m = 0$) is missing (or not detected), we use the average frequency of the $m = \pm 1$ components as central frequency. When one of the two side components ($m = \pm 1$) is missing (or not detected), we use the symmetric value with respect to that of $m = 0$ as its frequency. When the two side components (both $m = +1$ and $m = -1$) are missing (or not detected) the stellar spin inclination angle must be very low. In this latter case the choice of the frequencies of the split components does not typically influence the fit. However, we approximate these frequencies using a reasonable value of the average rotational splitting $\overline{\delta\nu}_\mathrm{rot}$, as obtained from the complete set of peak-bagged dipolar mixed modes of the star (see Supplementary Tables 1 and 2 for all the values of $\overline{\delta\nu}_\mathrm{rot}$).

Finally, we perform the fits using a Bayesian parameter estimation of the inclination angle $\theta$, which represents the free parameter of the 3-component model profile previously defined. By varying the inclination angle, the relative heights of the three $m$ components change by redistributing the total power of the multiplet among its different peaks. When the inclination angle is low ($\theta \approx 0^\circ$), the central component contains most of the power, while for high inclination angles ($\theta \approx 90^\circ$) the central component disappears because the power is equally distributed to the two side components, which both exhibit the same height. For the parameter estimation we use a uniform prior on $\cos \theta$ for $5^\circ < \theta < 85^\circ$ and a uniform prior on $\theta$ for $0^\circ \leq \theta \leq 5^\circ$ and $85^\circ \leq \theta \leq 90^\circ$, and an exponential likelihood function, as adopted for the fitting of the power spectra\cite{Corsaro14}. The multiplets are fit all together at the same time, for providing a more robust and better constrained estimate of $\theta$. An example of the resulting fit to $\theta$ for one red giant star of our sample is shown in Supplementary Fig.~1, while the inclination angles are given in Supplementary Tables 1 and 2. The same fitting technique is applied to the control sample, and the results are listed in Supplementary Table 5.

\noindent 
{\textbf{Simulation of three-dimensional distribution of uniform stellar-spin orientation.}} When observing a sample of stars with a uniform orientation of their stellar-spin vectors in a three-dimensional space, the geometrical effect from the line of sight shown in Fig. 1a, introduces an observational bias toward high inclination angles. This is happening because the number of stars per projected inclination angle is proportional to the associated solid angle, which follows the law $d\Omega = \sin \theta d\theta$. Therefore, to obtain a realistic distribution of a uniform orientation of the spin vector in three dimensions we simulate a uniform (flat) distribution in $-1 \leq \cos \theta \leq 1$, and subsequently fold it into $0 \leq \cos \theta \leq 1$ to account for the degeneracy between $\theta$ and ($\theta - \pi$) arising from the symmetry of the problem. This yields the theoretical histogram shown overlaid in Fig. 1b, c, and 3a, b.

\noindent
{\textbf{Probability of finding two clusters with strong stellar-spin alignment distributions.}} We first describe the distributions of spin inclinations shown in Fig. 1b, c using a Gaussian having the mean stellar-spin inclination angle for a cluster as a centroid $\overline{\theta}$, and a standard deviation $\sigma_{\theta}$ reflecting the sample dispersion around the mean. This yields an interval in inclination angle for each cluster defined as $\left[ \overline{\theta} - \sigma_{\theta}, \overline{\theta} + \sigma_{\theta} \right]$. The intervals are $\left[14^\circ, 39^\circ \right]$ and $\left[6^\circ,51^\circ \right]$ for NGC~6791 and NGC~6819, respectively. Then we consider a random distribution of spin-inclinations as described in the previous paragraph, and derive the cumulative probability of measuring an inclination angle in the range $\left[ \theta_1, \theta_2\right]$, with $\theta_1 < \theta_2$, to be $p = \int_{\theta_1}^{\theta_2} d\Omega$, which using the definition of solid angle described in the previous paragraph results in $p = \cos \theta_1 - \cos \theta_2$. For the inclination intervals defined by our distributions we obtain that $p_\mathrm{NGC~6791} \simeq 20\,\%$, and $p_\mathrm{NGC~6819} \simeq 36\,\%$, which yields the joint probability of $\sim7.2\%$ to find two clusters with the observed distributions of stellar-spin inclinations. While this value is low, one needs to realize that strong alignment is a more likely conclusion if the inclination angle is small. This is because a spin-inclination distribution of a strong alignment that peaks at low angles, stands out more significantly relative to the expected distribution from a uniform orientation, which instead has its maximum at $\theta = 90^\circ$ (see Fig. 1b, c, and 3a, b).

\noindent 
{\textbf{Three-dimensional hydrodynamical simulations of proto-cluster formation.}} The simulations of cluster formation are performed with RAMSES\cite{Teyssier02,Fromang06}, a magnetohydrodynamic (MHD) code with adaptive mesh refinement. We simulate the self-gravitating collapse of a molecular cloud following ideal MHD equations. We initialize the simulations with a Bonnor-Ebert-like spherical molecular cloud\cite{Lee16} of $10^3 M_{\odot}$ and density

\begin{equation*}
\rho(r) = \rho_0 \left[ 1 + \left( \frac{r}{r_0} \right)^2 \right]^{-1} \, .
\end{equation*}
A polytropic equation of state is used such that the gas is isothermal 10\,K at low density and heats up adiabatically as $T \propto \rho^{2/3}$ at a turn-over density of $10^{10}$ H$_2$\,cm$^{-3}$. The molecular cloud is seeded with an initial turbulent velocity field, solid body global rotation, and no magnetic field. We investigate initial conditions with different levels of turbulence and rotation. We list the parameters in Supplementary Table~3. 

The initial molecular cloud is set to be very compact and dense such that a mass of $10^3\,M_{\odot}$ is contained in a sphere having a diameter of 0.17\,pc. This choice was made due to the compromise between obtaining sufficient statistics --- for achieving a number of forming stars comparable to that from the observations --- and resolving the dynamics of the cloud at stellar scales\cite{Lee16}, namely a few astronomical units. Despite the initial size, the cluster expands after gas expulsion due to the shallowing of the gravitational potential.

To study the stars formed inside the molecular cloud, sink particle algorithms are used\cite{Lee16}. Once the density reaches the threshold of $10^{10}$\,H$_2$\,cm$^{-3}$, we detect the mass concentration and check whether the mass is virially unstable and the flow is locally converging. Subsequently, sink particles are used to follow the Lagrangian mass. The momentum and the angular momentum of the accreted material is added to the sink. We then consider the stars having accreted more than 0.7\,$M_\odot$ and examine their spin axis. Varying the mass threshold toward higher values has no impact on the significance of the alignment level. This mass threshold is thus a reasonable value above which stars begin to show a significant spin alignment.

With simulations, we have access to the three-dimensional information and we can choose to project the quantities in any selected orientation. We show that certain initial conditions result in a stellar cluster that has a preferred orientation of rotation. The choice of the projection vector $\vec{z}$ is as follows: we have the matrix $\bf{J}$ of which each of the $N$ rows is the unit vector $\vec{j}_i$ of a single star's spin axis, where $N$ is the number of stars considered. To find the maximal alignment, we maximize the quantity

\begin{equation*}
\sum_{i=1}^{N} \left( \vec{j}_i \cdot \vec{z} \right)^2 = \vec{z}^T \mbox{\bf J}^T \mbox{\bf J \,} \vec{z} \, .
\end{equation*}
The vector that gives us the best observed alignment is the eigenvector corresponding to the largest eigenvalue $\alpha$ of the matrix $\mbox{\bf J}^T\mbox{\bf J}/N$. If the orientation is uniform in three-dimensions, we expect this matrix to have three equal eigenvalues at $1/3$. 

\noindent 
{\textbf{Spin-alignment level and its significance.}} The alignment coefficient of a sample of $N$ stars is 
\begin{equation*}
\alpha = \frac{1}{N} \sum_{i=1}^N \cos^2 \theta_i \, .
\end{equation*}
For a perfect alignment along the line of sight $\alpha = 1$, while for $\alpha = 1/3$ the distribution of projected spin inclinations coincides with the ideal uniform distribution in three dimensions, obtained for $N \rightarrow \infty$. In the real case with a finite number of stars, the associated uniform distribution can show an intrinsic alignment ($\alpha > 1/3$). To quantify its significance according to the number of stars in our sample we have performed five sets of $10000$ simulations of three-dimensional uniform stellar-spin orientation, hence we computed the alignment coefficient for each set. The resulting values of $\alpha$ are listed in Supplementary Table~4, and they decrease with increasing number of stars. Using the statistical test of the p-value with the null hypothesis of having a three-dimensional uniform orientation of spin axes, these simulations show that for Simulation 1 (Supplementary Table 3) there is no alignment, for Simulations 2 and 3 only a marginal alignment (about $1\sigma$), and for Simulation 4, representing the case with the strongest rotational energy contribution, we have a significant alignment (over $3\sigma$). The observed distributions for NGC~6791 and NGC~6819 are largely significant (over $6$ and $5\sigma$, respectively). However, for the independent control sample of $36$ stars, the alignment coefficient is in agreement (within $1\sigma$) with the expected value for spin axes randomly distributed in three dimensions. The agreement (within $1\sigma$) with a random distribution is found even by considering subsamples of field stars having the same evolutionary stage.

\noindent
{\textbf{Comparison with an independent sample of stars in NGC~6819.}}
A sample of 30 main-sequence stars observed by Kepler and belonging to NGC~6819 has measurements of rotational periods ($P_\mathrm{rot}$), with a subsample of 25 stars having also projected equatorial velocities ($v\sin \theta$) available\cite{Meibom15}. Among the 10 stars having $P_\mathrm{rot} < 17$ days (hereafter fast rotating stars), the five stars in the range $5 < v\sin\theta < 15$ km s$^{-1}$ are cluster members that hint at high spin-inclination angles because they match $P_\mathrm{rot}$--$v \sin \theta$ tracks at constant stellar radius and $\theta = 90^\circ$. These stars may not appear compatible with the average low spin-inclination angle found for NGC~6819 in this work. However our reanalysis of the rotational periods for the entire sample of 30 stars reveals that, while confirming most of the published measurements for stars having $P_\mathrm{rot} > 17$ days, we cannot confirm the periods for the 10 fast rotating stars. We note that this has no implications on the gyrochronology results of NGC~6819 since all these fast rotating stars are the most massive of the sample and were not used for estimating the cluster age\cite{Meibom15}. 

The published rotation periods for the fast rotating stars were obtained using light curves produced by a version of the NASA Kepler data processing pipeline known to affect the stellar signal with periods longer than 3 days due to high-pass filtering\cite{Garcia13}; thus potentially biasing stars with slow rotation toward a false fast rotation. Hence for our measurements of rotation periods we instead use a different methodology to prepare light curves\cite{Garcia11,McQuillan14,Garcia14wavelet,Ceillier16} that has been shown to be robust also for slow rotators\cite{Aigrain15}. 

Among the 10 fast rotating stars we find reliable periods for only four stars (KIC~5111834, KIC~5023899, KIC~5023760, KIC~5113601). Their periods range from 25 to about 40 days, hence significantly longer than those already published. For these stars the amplitude of the modulation is small ($\sim1000$ parts-per-million magnitude). This could arise either because of star spots seen from a low spin-inclination angle, or blending with fainter cool active main-sequence stars. We conclude that for these stars further observations are required to confirm their published $v \sin \theta$ (available only for KIC~5111834 and KIC~5113601), and therefore to be able to disentangle the two mentioned effects. For the remaining six fast rotating stars we find that two stars (KIC~5024227 and KIC~5026583, with the latter having $v \sin \theta$ available) do show an ambiguous rotation period in our analysis and that four stars (KIC~4938993, KIC~5024122, KIC~5111207, KIC~5112499, with the first two having $v \sin \theta$ measurements) could be polluted by nearby stars seen in United Kingdom Infra-Red Telescope (UKIRT) high-resolution images. In the latter four cases, one or two additional stars are located within a four-Kepler-pixel square around the target, preventing us from being certain of which star is actually responsible for the alleged rotational signal found in the Kepler photometry. We therefore caution the use of these six stars when drawing conclusions about the general rotational properties of the cluster. The comparison is shown in Supplementary Fig. 2 and 3. In conclusion, about 70\% of the main-sequence cluster stars having both published $v \sin \theta$ and reliable period detections from our analysis, are compatible with a low spin-inclination regime, therefore in agreement with the findings presented in this work.

\noindent
{\textbf{Data availability.}} The datasets analyzed and the results generated during the current study are available in the CEA Saclay repository, \\ \href{http://irfu.cea.fr/Phocea/Vie\_des\_labos/Ast/ast\_technique.php?id\_ast=3801}{http://irfu.cea.fr/Phocea/Vie\_des\_labos/Ast/ast\_technique.php?id\_ast=3801}. The original raw Kepler light curves were obtained from the MAST, available at \href{https://archive.stsci.edu/index.html}{https://archive.stsci.edu/index.html}. The power spectra were obtained using the commercially available IDL routine {\it lnp\_test}. The inpainting interpolation software K-Inpainting is publicly available at \\
\href{http://irfu.cea.fr/Sap/en/Phocea/Vie_des_labos/Ast/ast_visu.php?id_ast=3346}{http://irfu.cea.fr/Sap/en/Phocea/Vie\_des\_labos/Ast/ast\_visu.php?id\_ast=3346}. The Bayesian inference code \diamonds\,\,is publicly available and documented at \\ \href{https://fys.kuleuven.be/ster/Software/Diamonds/}{https://fys.kuleuven.be/ster/Software/Diamonds/}. The simulation of the three-dimensional uniform orientation of the spin vectors and the Bayesian fits to the projected inclination angles are obtained using built-in IDL routines and subroutines, which are therefore not made publicly available. The RAMSES code package for the three-dimensional hydrodynamical simulations is publicly available and documented at \href{https://bitbucket.org/rteyssie/ramses/wiki/}{https://bitbucket.org/rteyssie/ramses/wiki/}.


\end{document}